\documentclass[twocolumn,showpacs,amsmath,amssymb]{revtex4}
\usepackage{bm}
\begin{document}
\newcommand{\pp}[1]{\phantom{#1}}
\newcommand{\be}{\begin{eqnarray}}
\newcommand{\ee}{\end{eqnarray}}
\newcommand{\ve}{\varepsilon}
\newcommand{\vs}{\varsigma}
\newcommand{\vp}{\varphi}
\newcommand{\Tr}{{\rm Tr\,}}

\title{
Relativistic BB84, relativistic errors, and how to correct them
}
\author{Marek Czachor and Marcin Wilczewski}
\affiliation{
Katedra  Fizyki Teoretycznej i Metod Matematycznych \\ 
Politechnika Gda\'{n}ska, ul. Narutowicza 11/12, 80-952 Gda\'{n}sk, Poland\\
E-mail: mczachor@sunrise.pg.gda.pl
}

\begin{abstract}
The Bennett-Brassard cryptographic scheme (BB84) needs two bases, at least one of them linearly polarized. The problem is that linear polarization formulated in terms of helicities is not a relativistically covariant notion: State which is linearly polarized in one reference frame becomes depolarized in another one. We show that a relativistically moving receiver of information should define linear polarization with respect to projection of Pauli-Lubanski's vector in a principal null direction of the Lorentz transformation which defines the motion, and not with respect to the helicity basis. Such qubits do not depolarize.
\end{abstract}
\pacs{PACS numbers: 03.67.Dd, 03.65.Ud}
\maketitle



In non-relativistic quantum mechanics a generic state of a free particle with spin 
takes the form where spin and momentum degrees of freedom are non-entangled, i.e. 
\be
\left(
\begin{array}{c}
\psi_{0}(\bm p)\\
\psi_{1}(\bm p)
\end{array}
\right)
&=&
\left(
\begin{array}{c}
\psi_{0}\\
\psi_{1}
\end{array}
\right)\psi(\bm p).\label{Gal}
\ee
This is the reason why it is possible to base the concept of a non-relativistic qubit on a 2-dimensional Hilbert space. In particular, observables asociated with spin are always of the form $A\otimes \bm 1$, where $\bm 1=\int d^3p|\bm p\rangle\langle\bm p|$ is the identity in momentum space and $A$ stands for a spin operator. The formula
$
\Tr\rho(A \otimes\bm 1)
=
\Tr_r \rho_r A.
$
allows to define states of qubits in terms of $2\times 2$ reduced density matrices. 

In relativistic quantum mechanics a generic state satisfies
\be
\left(
\begin{array}{c}
\psi_{0}(\bm p)\\
\psi_{1}(\bm p)
\end{array}
\right)
&\neq&
\left(
\begin{array}{c}
\psi_{0}\\
\psi_{1}
\end{array}
\right)\psi(\bm p).\label{rel}
\ee
The origin of this property is very deeply rooted in the structure of unitary representations of the Poincar\'e group. A qubit which in one reference frame takes the form (\ref{Gal}) will be seen in a form (\ref{rel}) by another observer. A Poincar\'e transformation necessarily involves multiplication by 
$\bm p$-dependent $SU(2)$ matrices, a fact making the form (\ref{Gal}) non-covariant. Definitions of qubits in terms of reduced density matrices with traced-out momenta are no longer justified. This is why quantum information theory based on such a formal notion of qubit \cite{PTS,PT,GA,BGA} is in danger of internal physical inconsistency. 

Constructing nonzero-spin unitary representations of the Poincar\'e group we always encounter certain spinor structure. The simplest representation corresponds to mass $m$ and spin 1/2. Whenever we write the state in a form (\ref{rel}) we 
{\it implicitly\/} choose a `spin quantization axis' and spin is here associated with the second Casimir invariant of the group, $W^aW_a$, where $W_a$ is the Pauli-Lubanski (PL) vector. 

The most popular choice of quantization axis corresponds to a timelike direction 
$t^a=(1,0,0,0)$. The resulting spin operator $t^aW_a$ is proportional to the helicity (in order to obtain directly the helicity one should choose 
$t^a=(1/|\bm p|,0,0,0)$). In application to quantum cryptography we need several different yes-no observables and helicity eigenstates are not sufficient. Natural candidates for such yes-no observables are projectors on linear combinations of opposite helicities, i.e. linear polarizations. The problem with linear polarizations defined in terms of helicities is that different momentum components undergo different $SU(2)$  transformations. In the photon case the $SU(2)$ transformations are diagonal and multiply opposite helicities by phase factors whose phases are opposite and momentum-dependent: A  wave packet which is linearly polarized in one reference frame becomes a combination of different linear polarizations in another reference frame and, hence, depolarized (cf. Sec. 2.5 in \cite{Weinberg}). An exception occurs for massless wave packets consisting excusively of parallel momenta since the Wigner phase factor is independent of 
$|\bm p|$.  Below we shall see that geometrically the effect is rooted in non-invariance of $t^a=(1,0,0,0)$ under nontrivial Lorentz boosts.

Different quantization axes lead to different yes-no observables. Taking 
$t^a=(0,\bm t)$ we arrive at observables equivalent to spins defined via relativistic center of mass \cite{Fleming} or, in the Dirac case, to the so-called even part of the Dirac spin. The even part is obtained from Dirac's $\bm \Sigma$ by sanwitching it between projectors on signs of energy. Physically one eliminates in this way the {\it Spinbewegung\/} oscillations \cite{Guertin-Guth}. The first applications of such spins to the relativistic EPR problem were given by one of us many years ago (even part of Dirac's spin in \cite{MC84}, relativistic center of mass, the PL vector, and even spin in \cite{PRA97,SPIE}). Quite recently the review \cite{T} discusses in the same context Dirac's and rest-frame spins, however the link of Dirac's $\bm\Sigma$ to the results of \cite{PRA97} was here overlooked. 

During the past year various relativistic aspects of EPR correlations were discussed in a series of detailed works. The intriguing paper \cite{R1} starts with a definition of spin in terms of a generator of rotations but taken from a  representation of the Poincar\'e group involving a non-standard clock synchronization. This seems to be  the first work where details related to spatial localization of measurements were taken into account, and the conclusion is that EPR correlations might in principle reveal a preferred reference frame. An approach starting {\it ab initio\/} from the level of quantum electrodynamics can be found in \cite{Pachos}; as opposed to the approach advocated in \cite{PTS,T} the momentum degrees of freedom are taken into account in measurements of spin and the loophole of the argument given in \cite{PTS} is not present. The roles of appropriate choices of spin measurements in EPR experiments were discussed in \cite{Tera1,Tera2}. The authors stress differences of their analysis with the one given in \cite{PRA97}, but the main conclusion (the degree of violation of the Bell inequality decreases with increasing velocities of the observers) remains unchanged. Similar conclusions can be found in \cite{Ahn}. Finally, a completely new direction of theoretical investigation was initiated by the work \cite{AM1}, and extended in \cite{Anderson,AM2}. The problem is what happens if the observers move noninertially. One expects here a host of new quantum-field-theoretic phenomena related to inequivalence of vacua in accelerated frames. 
The situation awaits a detailed review, especially in the context of possible experiments.

In the present paper we concentrate on the choice of qubits appropriate for cryptographic problems involving relativistically moving observers. At the level of first quantization we do not experience the subtleties related to the Unruh effect \cite{AM1,Anderson,AM2} and can in principle discuss also noninertial motions. Quantization in curved spaces and accelerated systems is still an open problem \cite{BirrelDavies}, so we prefer to concentrate on purely kinematic phenomena which can be naturally treated at the level of representations of the Poincar\'e group.


We define qubits by vectors from the mass-$m$ spin-1/2 unitary representation of the (covering space of) the Poincar\'e group. In momentum space the qubits are given by pairs of functions $\big(f_{\it 0}(p),f_{\it 1}(p)\big)$, 
$p^2=m^2$. It is essential that whenever one writes the pair  
$\big(f_{\it 0}(p),f_{\it 1}(p)\big)$ one implicitly chooses a basis of states coresponding to a projection of the PL vector in some direction 
$t^a\in \bm R^4$ in Minkowski space. This choice is implicitly present in the transformation properties of the qubit. 

The PL vector $W_a=P^b {^*S}_{ab}$ is a tensor operator \cite{BR} and under the action of the Poincar\'e group its projection in the direction $t^a$ gets transformed by 
\be
{\cal U}_{y,\Lambda}^{-1}t^aW_a {\cal U}_{y,\Lambda}= t^a\Lambda_{a}{^b}W_b
\ee
where ${\cal U}_{y,\Lambda}$ is a unitary representation.
One can say that a moving particle experiences measurements of spins in Lorentz-modified directions $t^a\Lambda_{a}{^b}$. Of particular interest are the directions $t^a$ satisfying the eigenvalue condition 
$t^a\Lambda_{a}{^b}\sim t^b$ since they lead to Lorentz-invariant yes-no observables. Eigenvectors of Lorentz transformations are known to be given by {\it null\/} vectors ($t^2=t^at_a=0$) and any $\Lambda\in SL(2,C)$ possesses at least one and at most two such eigen-directions (principal null directions --- PNDs) 
\cite{Penrose}. 

Accordingly, it is very natural to contemplate projections of the PL vector in null directions instead of the usual timelike or spacelike ones.
Moreover, the projection of the PL vector in momentum direction vanishes, 
$P^aW_a=0$, and therefore we obtain a kind of gauge freedom: For any parameter 
$\theta$  the observables $t^aW_a$ and $(t^a+\theta P^a)W_a$ are identical. An appropriate choice of $\theta$ and a $P$-dependent $t^a$ will allow us to work with invariant yes-no observables which are equivalent to projections of PL vector in directions perpendicular to the 4-momentum. In what follows we shall describe a simple procedure leading to such invariant yes-no observables. 

A formalism which almost ideally suits the purposes of relativistic quantum information theory is the 2-spinor calculus, especially in the form developed by Penrose \cite{Penrose}. The unitary representations of the Poincar\'e group can be translated into a 2-spinor language by means of `Bargmann-Wigner (BW) spinors' \cite{MC-BW}. One exploits here the special role played in 2-spinor formalism by null directions. The 4-momentum $p_a$ is split into a linear combination of two null directions $\pi_a$ and $\omega_a$ defined by 
\be
p_a=\pi_a+(m^2/2)\omega_a=\pi_A\bar \pi_{A'}+(m^2/2)\omega_A\bar \omega_{A'}
\label{p}
\ee
where $\pi_A$, $\omega_A$ is a field of spin-frames, i.e. $\omega_A\pi^A=1$. If 
$\nu_A$ is any $p$-independent spinor, the spin frame may be taken as
\be
\omega^{A}(p)
&=&
\frac{\nu^{A}}{
\sqrt{p^{BB'}\nu_B
\bar \nu_{B'}}}
=
\omega^{A}(\nu,p),\label{omega}\\
\pi^A(p)
&=&
\frac{p^{AA'}\bar \nu_{A'}}
{\sqrt{p^{BB'}\nu_B
\bar \nu_{B'}}}
=
\pi^A(\nu, p).\label{pi}
\ee
One can directly verify that the spin-frame satisfies (\ref{p}) and
\be
\Lambda\pi_A(\nu, p)&=&\Lambda_{A}{^B}\pi_B(\nu, \Lambda^{-1}p)
=\pi_A(\Lambda\nu, p),\label{pi'}\\
\Lambda\omega_A(\nu, p)&=&\Lambda_{A}{^B}\omega_B(\nu, \Lambda^{-1}p)
=\omega_A(\Lambda\nu, p).\label{omega'}
\ee
The latter formulas will be crucial for our analysis of relativistic qubits. 
The simplest unitary representation is characterized by mass $m$ and spin-1/2 and its BW-spinor form reads
\be
{\cal U}_{y,\Lambda}f_{\cal A}(p)
=e^{iy\cdot p}U(\Lambda,p){_{\cal A}}{^{\cal B}}f_{\cal B}(\Lambda^{-1}p)
\label{U}
\ee
where $U(\Lambda,p){_{\cal A}}{^{\cal B}}\in SU(2)$, $p^2=m^2$. The BW-spinor indices are written in the calligraphic font to distinguish them from the 
$SL(2,C)$ ones. The matrix 
\be
U(\Lambda,p){_{\cal A}}{^{\cal B}}
&=&
\left(
\begin{array}{cc}
\omega{_{A}}(p)\Lambda\pi{^{A}}(p) & 
-\frac{m}{\sqrt{2}}\omega{_{A}}(p)
\Lambda\omega{^{A}}(p)\\
\frac{m}{\sqrt{2}}
{\bar \omega}{_{A'}}(p)\overline{\Lambda\omega}{^{A'}}(p) &
{\bar \omega}{_{A'}}(p)\overline{\Lambda\pi}{^{A'}}(p)
\end{array}
\right)\nonumber\\
\label{macierz}
\ee
is responsible for changes of the `polarization' whereas the part 
$f_{\cal B}(\Lambda^{-1}p)$ introduces Doppler shifts. 

One can show \cite{MC-BW} that the amplitudes $f_{\it 0}(p)$, $f_{\it 1}(p)$ play for Dirac electrons the roles of momentum-space wave functions associated with eigenvectors of the projection of the PL vector in the null direction 
$t^a=\omega^a$, with eigenvalues $-1/2$ and $+1/2$, respectively. Choosing $\theta=-m^{-2}$ we find that the projection of $\omega^aW_a$ is an observable  identical to the projection of $W_a$ in the direction $\omega^a-m^{-2}p^a$ 
which is spacelike and orthogonal to the 4-momentum. 

At the level of BW-spinors we do not have to make any reference to the Dirac equation but can directly compute the generators, the PL vector, and its projection in any direction. In particular, in momentum space the projection in the null direction turns out to be
\be
\omega^a(p) W_a(p){_{\cal A}}{^{\cal B}}=
\frac{1}{2}
\left(
\begin{array}{cc}
-1 & 0\\
0 & 1
\end{array}
\right)
=
-\frac{1}{2}\sigma_3
\label{sigma_3}
\ee
which agrees with the fact that the amplitudes $f_{\it 0}(p)$, $f_{\it 1}(p)$ correspond, respectively, to the eigenvalues $-1/2$ and $+1/2$ at the level of the Dirac equation. 


The spin operator we have introduced through projection of the PL vector on the null quantization axis $\omega^a(p)$ has led to the standard-looking spin. So what have we gained with respect to the earlier works where 
$\sigma_3$ was taken for granted as the correct operator associated with relativistic qubits? The gain is that we have arrived at $\sigma_3$ by means of a systematic procedure and have the relativistic transformation properties of qubits under control. Recall that in addition to (\ref{sigma_3}) we have the representation 
(\ref{U}) where the matrix $U(\Lambda,p){_{\cal A}}{^{\cal B}}$ does not, in general, commute with $\sigma_3$. 

Notice, however, that the actual problem we will need to solve in practical quantum communication is how to correct the errors which are due to a relativistic  and perhaps noninertial motion $s\mapsto\Lambda(s)$ of an observer. The problem can be reduced to an appropriate choice of quantization axis which defines the qubit. 

A PND associated with $\Lambda\in SL(2,C)$ is the flagpole direction of an eigenspinor of $\Lambda$, i.e. 
\be
\Lambda_{A}{^B}\nu_B=\lambda \nu_A\label{PND}
\ee
where $\lambda=|\lambda|e^{i\varphi}$ is in general complex.
Classification of PNDs of $SL(2,C)$ transformations can be found in \cite{Penrose}. Inserting (\ref{PND}) into  (\ref{omega})--(\ref{omega'}) we find 
\be
\Lambda\pi_A(\nu, p)= e^{-i\varphi}\pi_A(\nu, p),\quad
\Lambda\omega_A(\nu, p)=e^{i\varphi}\omega_A(\nu, p),\nonumber
\ee
and 
\be
U(\Lambda,p){_{\cal A}}{^{\cal B}}
&=&
\left(
\begin{array}{cc}
e^{-i\varphi} & 0\\
0 & e^{i\varphi} 
\end{array}
\right)\label{Uphi}
\ee
where $\varphi$ is momentum-independent. The independence of momentum is important since the transformation $s\to\Lambda(s)$ affects all the momentum components in the same way. 
An arbitrary linearly polarized state is now transformed as follows
\be
\left(
\begin{array}{c}
f_{\it 0}(p)\\
f_{\it 1}(p) 
\end{array}
\right)
\mapsto
\left(
\begin{array}{c}
{\cal U}_{y,\Lambda}f_{\it 0}(p)\\
{\cal U}_{y,\Lambda}f_{\it 1}(p) 
\end{array}
\right)
=
e^{iy\cdot p}
\left(
\begin{array}{c}
e^{-i\varphi}f_{\it 0}(\Lambda^{-1}p)\\
e^{i\varphi}f_{\it 1}(\Lambda^{-1}p) 
\end{array}
\right)\nonumber
\ee
and a linear polarization goes into linear polarization, perhaps rotated by some angle. In particular, the product form (\ref{Gal}) is covariant. This was possible only because we replaced helicity qubits by qubits related to an invariant direction. 

For quantum cryptographic protocols, such as BB84,  it may be important to allow for motions which are characterized by two different PNDs. 
As an example consider a general accelerated motion (of Alice, say) in a $z$-direction. The relevant $SL(2,C)$ transformation reads
\be
\Lambda{_{A}}{^{B}}
&=&
\left(
\begin{array}{cc}
w^{1/2} & 0\\
0 & w^{-1/2} 
\end{array}
\right),\nonumber
\ee
$w=\sqrt{1+\beta}/\sqrt{1-\beta}$, $\beta=v(s)/c$. The eigenvalues of $\Lambda$ are real and there are two eigenvectors
\be
\nu^{(-)}{_{A}}
=
\left(
\begin{array}{c}
1\\
0  
\end{array}
\right),\quad
\nu^{(+)}{_{A}}
=
\left(
\begin{array}{c}
0\\
1  
\end{array}
\right).\label{mu-nu}
\ee
In both cases we find $U(\Lambda,p)=\bm 1$ (since $\varphi=0$) and
\be
\left(
\begin{array}{c}
{\cal U}_{y,\Lambda}f_{\it 0}(\nu^{(\pm)},p)\\
{\cal U}_{y,\Lambda}f_{\it 1}(\nu^{(\pm)},p) 
\end{array}
\right)
&=&
e^{iy\cdot p}
\left(
\begin{array}{c}
f_{\it 0}(\nu^{(\pm)},\Lambda^{-1}p)\\
f_{\it 1}(\nu^{(\pm)},\Lambda^{-1}p) 
\end{array}
\right)\label{f-nu}
\ee
The amplitudes $f_{\cal A}(\nu^{(-)},p)$, $f_{\cal A}(\nu^{(+)},p)$ represent wave functions associated with projections of the PL vector in invariant directions 
$
\omega^a(\nu^{(\pm)},p)=(p^0\mp p^3)^{-1}(1,0,0,\pm 1).
$
The $p$-dependent denominators come from the denominator in 
$\omega_a(\nu^{(\pm)},p)=\nu^{(\pm)}_a/(p\cdot \nu^{(\pm)})$ and could be also skipped since the null directions of $\nu^{(\pm)}_a$ and $\omega_a(\nu^{(\pm)},p)$ are identical. The yes-no observables are defined by normalization of eigenvalues to $\pm 1$. This is similar to the problem of choosing $t^a$ associated with the helicity.

The formula (\ref{f-nu}) illustrates the role of appropriate choices of quantization axes. No rotations of qubits are involved, the relativistic corrections are reduced to the Doppler shifts, and the form (\ref{Gal}) will be preserved. 

The associated yes-no observables are given in both cases by $\sigma_3$ but, of course, two different bases are involved. For future references we give 
here the general form of the $SU(2)$ transformation which maps qubits associated with a null $\omega^a$ direction into those associated with a most general direction $t^a$ (null, timelike, or spacelike, and in general $p$-dependent). 
Let 
\begin{widetext}
\be
\left(
\begin{array}{c}
{\Omega}(t,p)_A\\
{\Omega}(t,p)_{A'}
\end{array}
\right)=-
\Big[8\lambda(t,p)
\bigl(
\lambda(t,p)
+
t\cdot p
-
m^2
(t\cdot\omega)
\bigr)\Big]^{-\frac{1}{2}}
\left(
\begin{array}{c}
(2\lambda(t,p)+t\cdot p) 
\pi{_{A}}
-
\pi{_{C}}t{^C}{_{X'}}p{_{A}}{^{X'}} 
+
3\frac{m^2}{2}t{_{AX'}}\bar \omega{^{X'}}
\\
\frac{m}{\sqrt{2}}
\Bigl(
(2\lambda(t,p)-t\cdot p) 
\bar \omega{_{A'}}
-
\bar \omega{_{C'}}t{_X}{^{C'}}p{^{X}}{_{A'}} 
-
3t{_{XA'}}\pi{^{X}}\Bigr) 
\end{array}
\right).\nonumber
\ee
\end{widetext}
Then
$
f_{\cal A}(t,p)
=
{\cal W}(p)_{\cal A}{^{\cal B}}
f_{\cal B}(\omega,p),
$
where
\be
{\cal W}(p)_{\cal A}{^{\cal B}}
=
\left(
\begin{array}{cc}\bar \omega(p)^{A'}
\bar {\Omega}(t,p)_{A'} &
\omega(p)^{A}
\bar {\Omega}(t,p)_{A}\\
-\bar \omega(p)^{A'}
{\Omega}(t,p)_{A'} &
\omega(p)^{A}
{\Omega}(t,p)_{A}
\end{array}
\right),
\ee
${\bar\Omega}(t,p)_{A}=\overline{{\Omega}(t,p)_{A'}}$, ${\bar \Omega}(t,p)_{A'}=
\overline{{\Omega}(t,p)_{A}}$, and $\lambda(t,p)=\sqrt{(t\cdot p)^2-m^2t^2}$. Notice that in momentum space the eigenvalues of the projection of the PL vector in a direction $t^a$ are given by $\pm\frac{1}{2}\lambda(t,p)$. 

Now, set $\omega^A(p)=\omega^A(\nu^{(-)},p)$, $\pi^A(p)=\pi^A(\nu^{(-)},p)$, and 
$t^a(p)=\omega^a(\nu^{(+)},p)$, where the invariant spinors are given by (\ref{mu-nu}). We find 
\be
f_{\cal A}(\nu^{(+)},p)
&=&
{\cal W}(p)_{\cal A}{^{\cal B}}
f_{\cal B}(\nu^{(-)},p),\label{BB84}\\
{\cal W}(p)_{\cal A}{^{\cal B}}
&=&
\frac{1}{\sqrt{1+|\zeta|^2}}
\left(
\begin{array}{cc}
|\zeta| & e^{i\chi}\\
-e^{-i\chi} & |\zeta|
\end{array}
\right)\in SU(2)\nonumber
\ee
where $\zeta=(p_1+ip_2)/m=|\zeta|e^{i\chi}$ is invariant under Lorentz boosts along the third axis. The matrix in (\ref{BB84}) itself is therefore also invariant under transformations which do not change the quantization axes.  

As a next application let us consider the case where the $SL(2,C)$ transformations do not commute with one another if taken at different points on the curve $s\mapsto \Lambda(s)$, i.e. 
$[\Lambda(s),\Lambda(s')]\neq 0$. A good illustration is a composition of the previously discussed boost with a null rotation i.e. 
\be
\Lambda(t){_{A}}{^{B}}
&=&
\left(
\begin{array}{cc}
w(s)^{1/2} & 0\\
0 & w(s)^{-1/2} 
\end{array}
\right)
\left(
\begin{array}{cc}
1 & \alpha(s)\\
0 & 1 
\end{array}
\right).
\ee
There is only one eigenvector, namely $\nu^{(-)}_A$. The construction is unchanged but we must make sure the spin is projected on the null axis $\omega^a(\nu^{(-)},p)$. This presence of only one invariant direction is not a problem since one direction is enough to define linear polarizations. 

Finally, let us consider the case of photons. The Poincar\'e transformation reads 
\be
{\cal U}_{y,\Lambda}f_{\cal AA}(p)
=e^{iy\cdot p}U(\Lambda,p){_{\cal A}}{^{\cal B}}U(\Lambda,p){_{\cal A}}{^{\cal B}}
f_{\cal BB}(\Lambda^{-1}p)
\label{U0}
\ee
where 
$
U(\Lambda,p){_{\cal A}}{^{\cal B}}
$
is the zero-mass version of (\ref{macierz}). There are only two degrees of freedom represented by $f_{\it 00}(p)$ and $f_{\it 11}(p)$. 

The diagonal elements
$U(\Lambda,p){_{\it 0}}{^{\it 0}}$, $U(\Lambda,p){_{\it 1}}{^{\it 1}}$ are momentum dependent phase factors. 
As before, an appropriate choice of $\nu$ in spin frames reduces $U(\Lambda,p)$ to
(\ref{Uphi})
where $\varphi$ is momentum independent. In the massless case there exists a topological restriction which does not occur for massive fields: For a fixed 
$\nu_A$ there exists a momentum $p_a$ parallel to the null vector 
$\nu_a=\nu_A\bar\nu_{A'}$ where the spin-frames are not defined  
(actually this happens for the direction given by the class of momenta parallel to $p_a$). Usually this is not a difficulty since one can define the spin-frames locally. In our problem this {\it is\/} a difficulty which restricts the choice of $\nu_a$ to null vectors which are not proportional to any 4-momentum in the wave packet. The case where the wave packet does contain a vector parallel to $\nu_a$ has to be treated with some care. 

The noise due to relativistic helicity-basis depolarization can be reduced 
for massless fields if the wave packets 
$f_{\cal AA}(p)$ consist of vectors parallel to some given direction. To what extent an approximation of a realistic signal by such a plane-wave-like front is acceptable depends on the geometry of the experiment. Qubits associated with PNDs do not impose any restriction on momentum-space wave packets.

We are grateful to J. Rembieli\'nski for comments.

\end{document}